\documentstyle[aps,preprint,12pt]{revtex}
\tighten
\title{Jost function for coupled channels
\thanks{talk given at XVI European Conference on Few--Body Problems in
Physics, Autrans, June 1998. To appear in Few Body Systems Suppl.
}
}
\author{S. A. Rakityansky
\thanks{{\it E-mail address:} rakitsa@kiaat.unisa.ac.za}
and S. A. Sofianos
\thanks{{\it E-mail address:} sofiasa@kiaat.unisa.ac.za}}
\address{Department of Physics, University of South Africa, P.O.Box 392,
Pretoria 0003, South Africa}
\begin{document}
\maketitle
\bigskip
Almost any textbook on the scattering theory has a chapter devoted to the
Jost function, but none of them gives a practical recipe for its calculation.
Thus  one usually gets a feeling that the Jost function is a pure
mathematical object, elegant and useful in formal theory, but impractical in
computations. This is even more so in the case of multi-channel problems,
where the Jost function is, in fact, a matrix. Recently, we proposed an exact
method for direct calculation of the Jost function for central
potentials (which may have Coulombic tails) \cite{nc,jpa1} and the Jost
matrix for non-central short range potentials \cite{jpa2}. This method works
for all real or complex momenta of physical interest, including 
the spectral points corresponding to bound and Siegert states. In this 
work we extend it for potentials which couple channels  with different 
thresholds.

There are three different types of physical problems associated with the
Schr\"odinger equation: bound, scattering, and resonance state problems. They
differ in the boundary conditions imposed on the wave function at large
distances. Alternatively, a solution can be specified by the boundary
conditions at the origin. For example, in the single--channel case 
the solution which vanishes as $r\to 0$ exactly like the Riccati-Bessel 
function, is called the {\it regular} solution. It is well known that if the
potential couples $N$ channels, the Schr\"odinger equation
\begin{equation}
\label{shcr}
\left[\partial^2_r+2\mu_n(E-E_n)-\frac{\ell(\ell+1)}{r^2}\right]
	       u_n(E,r)=\sum_{n'=1}^NV_{nn'}(r)u_{n'}(E,r)
\end{equation}
has $2N$ linearly independent column--solutions
$$
		   \left(
		   \begin{array}{c}
		   u_1\\
		   u_2\\
		   \vdots\\
		   u_N\\
		   \end{array}
		   \right)_{\!\!\!1}\!,
		   \,\,\left(
		   \begin{array}{c}
		   u_1\\
		   u_2\\
		   \vdots\\
		   u_N\\
		   \end{array}
		   \right)_{\!\!\!2}\!,
		   \,\,\left(
		   \begin{array}{c}
		   u_1\\
		   u_2\\
		   \vdots\\
		   u_N\\
		   \end{array}
		   \right)_{\!\!\!3}\!,\cdots
		   \,\,\left(
		   \begin{array}{c}
		   u_1\\
		   u_2\\
		   \vdots\\
		   u_N\\
		   \end{array}
		   \right)_{\!\!\!2N}
$$
and only half of them are regular at the point $r=0$. We can therefore
combine all the regular solutions (to distinguish them we use the notation
$\phi$) in a square matrix,
$$
		   \left(\!\!\!
		   \begin{array}{c}
		   \phi_1\\
		   \phi_2\\
		   \!\!\vdots\\
		   \phi_N\\
		   \end{array}
		   \!\!\!\right)_{\!\!\!1}\oplus
		   \left(\!\!\!
		   \begin{array}{c}
		   \phi_1\\
		   \phi_2\\
		   \!\!\vdots\\
		   \phi_N\\
		   \end{array}
		   \!\!\!\right)_{\!\!\!2}\oplus\cdots
		   \raisebox{-3pt}{\mbox{\Huge{$\Rightarrow$}}}
		   \left(\!
		   \begin{array}{cccc}
		   \!\!\phi_{11} & \phi_{12} & \cdots & \phi_{1N}\\
		   \!\!\phi_{21} & \phi_{22} & \cdots & \phi_{2N}\\
		   \!\!\!\!\vdots& \!\!\vdots& \ddots & \!\!\vdots\\
		   \!\!\phi_{N1} & \phi_{N2} & \cdots & \phi_{NN}\\
		   \end{array}
		   \!\!\!\right) \equiv \|\phi(E,r)\|\ ,
$$
which obeys the (matrix) Schr\"odinger equation with the boundary 
condition
\begin{equation}
\label{bcon}
	\lim\limits_{r\to 0}\displaystyle
	\phi_{nn'}(E,r)/\displaystyle j_{\ell}(k_{n'}r)=\delta_{nn'}
\end{equation}
where $j_\ell(z)$ is the Riccati--Bessel function and
$k_n=\sqrt{2\mu_n(E-E_n)}$ is the channel momentum. Any physical solution
must be regular at the origin and therefore it is  a linear 
combination of the regular columns. In this respect the
matrix $\|\phi\|$  can be  called the {\it regular basis}.

In order to obtain this basis in the form which is most suitable for
describing its long--range asymptotics, we replace the  unknown
matrix--function $\|\phi(E,r)\|$ by
two others, $\|{\cal F}^{(-)}(E,r)\|$ and $\|{\cal F}^{(+)}(E,r)\|$,
via the following ansatz
\begin{equation}
\label{ansatz}
   \phi_{nn^\prime}(E,r) = \displaystyle{\frac12}\left[h_{\ell}^{(-)}(k_nr)
   {\cal F}_{nn'}^{(-)}(E,r)+h_{\ell}^{(+)}(k_nr)
   {\cal F}_{nn'}^{(+)}(E,r)\right]\ ,
\end{equation}
with the additional constraint conditions
\begin{equation}
\label{lagrange}
	h_\ell^{(-)}(k_nr)\partial_r{\cal F}_{nn'}^{(-)}(E,r)+
	h_\ell^{(+)}(k_nr)\partial_r{\cal F}_{nn'}^{(+)}(E,r)=0
\end{equation}
which make the derivative of the wave functions continuous even if the
potential has a sharp cut--off. Substituting this ansatz into the
Schr\"odinger equation we derive the following system of first 
order differential equations for these new unknown matrices, 
\begin{equation}
\label{difeq}
     \partial_r{\cal F}_{nn'}^{(\mp)} =
     \mp\displaystyle\frac{1}{2ik_n}
     h_{\ell}^{(\pm)}(k_nr)\sum_{n''}V_{nn''}(r)
     \left[h_{\ell}^{(+)}(k_{n''}r){\cal F}_{n''n'}^{(+)}+
     h_{\ell}^{(-)}(k_{n''}r){\cal F}_{n''n'}^{(-)}\right]\ .
\end{equation}
The boundary conditions corresponding to (\ref{bcon}) are
\begin{equation}
\label{bconf}
	\lim\limits_{r\to 0}\displaystyle
	{\cal F}_{nn'}^{(\pm)}(E,r)=\delta_{nn'}\ .
\end{equation}
Explicit inclusion of the Riccati--Hankel functions into the basis
(\ref{ansatz}) guarantees the correct asymptotic behaviour of the physical
solution. 

In constructing a physical wave function $\varphi_n(E,r)$ one
requires the appropriate coefficients $c_n$ in the
linear combination of the regular columns,
$$
   \varphi_n(E,r)=\sum_{n'}\phi_{nn'}(E,r)c_{n'}\ .
$$
For bound and resonant states, for example, 
in the physical wave function  only the term proportional to $h^{(+)}$
survives at large distances. This gives the  homogeneous system
of equations
$$
	\sum\limits_{n'}{\cal F}_{nn'}^{(-)}(E,\infty)c_{n'}=0
$$
which has a non-trivial solution if and only if its determinant
is zero, i. e.,
\begin{equation}
\label{det0}
	\det\|{\cal F}^{(-)}(E,\infty)\|=0\ .
\end{equation}
It is easily seen that this is the determinant of the Jost matrix $\|F(E)\|$
which is defined by the asymptotic behaviour of the regular basis
 \cite{taylor}
$$
   \phi_{nn^\prime}(E,r) \mathop{\longrightarrow}\limits_{r\to\infty}
   \displaystyle{\frac12}\left[h_{\ell}^{(-)}(k_nr)
   F_{nn'}(E)+h_{\ell}^{(+)}(k_nr)
   \tilde{F}_{nn'}(E)\right]\ .
$$
We can calculate this matrix by solving the differential equations 
(\ref{difeq})
from $r=0$ up to a large radius where the potential is insignificant. Since
the potential vanishes, the right hand sides of Eqs.(\ref{difeq}) disappear
at large distances. Hence the derivatives
$\partial_r\|{\cal F}^{(\mp)}(E,r)\|$ become zero which means that
asymptotically the matrix--functions $\|{\cal F}^{(\mp)}(E,r)\|$ become
constants coinciding with $\|F(E)\|$ and $\|\tilde{F}(E)\|$ respectively.

Similarly to the single--channel case (see Refs. \cite{nc,jpa1,jpa2}) 
the above scheme can be easily implemented only for 
bound and scattering states. In the resonance domain
of the complex energy plane the Riccati--Hankel functions
$h_\ell^{(+)}(k_nr)$, and therefore the derivatives of
$\|{\cal{F}}^{(-)}(E,r)\|$ at large distances,  become infinitely large which
means that the limit $\|{\cal F}^{(-)}(E,\infty)\|$ does not exist. Thus
in order to calculate the Jost matrix in this domain we need to perform
a complex rotation of the independent variable  $r$,
$r=x\exp(i\theta)$,  in Eqs.(\ref{difeq}), where $x\ge 0$ and
$0\le\theta<\pi/2$. This results in a movement of the unitary cuts at 
each threshold downwards as is shown in Fig. 1. After such a rotation
$h_\ell^{(+)}(k_nr)$ vanishes in all areas above the turned cuts, which
guarantees the existence of the limit $\|{\cal F}^{(-)}(E,\infty)\|$ there.

\bigskip
\unitlength=0.3mm
\begin{picture}(230,100)
\put(250,40){\parbox{7cm}{
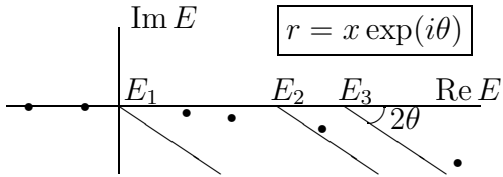
Figure 1. \it Complex rotation of the coordinate by an angle $\theta$
results in a  rotation of the unitary cuts by $2\theta$.}}
\put(60,40){%
\begin{picture}(0,0)%
\put(-50,0){\line(1,0){210}}
\put(0,-30){\line(0,1){65}}
\put(70,30){\fbox{$r=x \exp(i\theta)$}}
\put(0,0){\line(3,-2){45}}
\put(70,0){\line(3,-2){45}}
\put(100,0){\line(3,-2){45}}
\put(110,0){\oval(15,15)[br]}
\put(120,-10){$2\theta$}
\put(2,3){$E_1$}
\put(67,3){$E_2$}
\put(97,3){$E_3$}
\put(5,35){${\rm Im\,}E$}
\put(140,3){${\rm Re\,}E$}
\put(-40,0){\circle*{4}}
\put(-15,0){\circle*{4}}
\put(30,-3){\circle*{4}}
\put(50,-5){\circle*{4}}
\put(90,-10){\circle*{4}}
\put(150,-25){\circle*{4}}
\end{picture}
}
\end{picture}
\bigskip

By locating the zeros of the Jost--matrix determinant we can find the
resonance energies and the total widths. 
The determination of the partial widths, however, is a difficult task 
for most methods used in the past. 
In our approach this is not a problem at all. Indeed, knowing
the Jost matrix, we can easily obtain the $S$--matrix at a given complex
energy as
$$
     \|S(E)\|=
     \|{\cal F}^{(+)}(E,\infty)\|\cdot\|{\cal F}^{(-)}(E,\infty)\|^{-1}
$$
which, generally, consists of two terms, viz.
$$
     S_{nn'}(E)=S_{nn'}^{{\rm bg}}-
     i\frac{\sqrt{\Gamma_n\Gamma_{n'}}}{E-E_r+i\Gamma/2}\ .
$$
One of them is the smooth background term  $S_{nn'}^{{\rm bg}}$
and the other a singular resonance term. Thus a partial width is the limit
$$
     \Gamma_n=
     \left|\lim_{E\to (E_r-i\Gamma/2)}(E-E_r+i\Gamma/2)S_{nn}(E)\right|\ .
$$
Practically, to find such a limit we calculate the $S$--matrix at a point 
very close to the resonant energy and multiply it by the corresponding 
energy difference.

To demonstrate the ability of the method we consider a numerical example
proposed by Noro and Taylor in Ref.\cite{NT} where they used the
$S$--wave potential
$$
	   \|V(r)\|=\left(
	   \begin{array}{rr}
	   -1.0 & -7.5\\
	   -7.5 & 7.5 \\
	   \end{array}
	   \right)r^2e^{-r}
$$
which describes a two--channel system with the thresholds $E_1=0$ and
$E_2=0.1$ (atomic units). By expanding the resonance wave function in a
series of square--integrable functions, Noro and Taylor located the first
resonance generated by this potential. We reproduce their result and found
also three bound states and several broad resonances. For comparison we put
the parameters of the first resonance in Table 1. The digits shown  do
not change when the rotation angle is changed. This demonstrates  the
accuracy of our method.
\begin{center}
\begin{tabular}{|c|c|c|l}
\cline{1-3}
 & our results & Noro $\&$ Taylor    &\\
\cline{1-3}
$E_r$      & 4.7681968188 & 4.7682   & \qquad Table 1.\\
$\Gamma$   & 0.0014201920 & 0.001420 & \qquad {\it Values are given
					       in atomic units}\\
$\Gamma_1$ & 0.000051103 & 0.000059  &\\
$\Gamma_2$ & 0.001368733 & 0.001361  &\\
\cline{1-3}
\end{tabular}
\end{center}

In conclusion, we  propose a method  based on simple 
differential equations of the first order, which can be easily solved
numerically. Thus, the spectrum generated by any given potential can be
thoroughly investigated.  At the same time the physical wave function 
can be obtained having the correct asymptotic behaviour.

\smallskip


\makeatletter \if@amssymbols%
\else\relax\fi\makeatother

\end{document}